\documentclass[graybox]{svmult}

\usepackage{mathptmx}       
\usepackage{helvet}         
\usepackage{courier}        
\usepackage{type1cm}        
\usepackage{makeidx}         
\usepackage{graphicx}        
\usepackage{multicol}        
\usepackage[bottom]{footmisc}

\makeindex             

\begin{document}

\title*{Coupled-Channel Effects in Collisions between Heavy Ions near the
Coulomb Barrier}
\author{C. Beck}
\institute{C. Beck  \at Institut pluridisciplinaire Hubert Curien, 
IN$_{2}$P$_{3}$-CNRS and Universit\'e de Strabourg - 23, rue du Loess BP 28,
F-67037 Strasbourg Cedex 2, France,
\email{christian.beck@iphc.cnrs.fr}
}
%
%
\maketitle

\abstract{With the recent availability of state-of-the-art 
heavy-ion stable and radioactive beams, there has been a renew 
interest in the investigation of nuclear reactions with heavy 
ions. I first present the role of inelastic and transfer 
channel couplings in fusion reactions induced by stable
heavy ions. Analysis of experimental fusion cross sections by
using standard coupled-channel calculations is discussed.
The role of multi-neutron transfer is investigated in
the fusion process below the Coulomb barrier by analyzing 
$^{32}$S+$^{90,96}$Zr as benchmark reactions. The enhancement 
of fusion cross sections for $^{32}$S+$^{96}$Zr is well 
reproduced at sub-barrier energies by NTFus code calculations 
including the coupling of the neutron-transfer channels 
following the Zagrebaev semi-classical model. Similar 
effects for $^{40}$Ca+$^{90}$Zr and $^{40}$Ca+$^{96}$Zr fusion 
excitation functions are found. The breakup coupling in both the 
elastic scattering and in the fusion process induced by weakly 
bound stable projectiles is also shown to be crucial. In this
lecture,
full coupled-channel calculations of the fusion excitation
functions are performed by using the breakup coupling 
for the more neutron-rich reaction and for the more weakly 
bound projectiles. I clearly demonstrate that 
Continuum-Discretized Coupled-Channel calculations are capable 
to reproduce the fusion enhancement from the breakup coupling 
in $^{6}$Li+$^{59}$Co.}

\section{ Introduction}

Heavy-ion fusion reactions at bombarding energies at the vicinity 
and below the Coulomb barrier have been widely studied 
\cite{Balantekin,Dasgupta,Liang,Canto,Keeley}. In low-energy 
fusion reactions, the very simple one-dimensional barrier-penetration 
model (1D-BPM) \cite{Balantekin,Dasgupta} is based upon a real 
potential barrier resulting from the attractive nuclear and repulsive 
Coulomb interactions. For light- and medium-mass nuclei, one only 
assumes that the di-nuclear system (DNS) fuses as soon as it has reached 
the region inside the barrier i.e. within the potential pocket. 
If the system can evolve with a bombarding energy high enough to 
pass through the barrier and to reach this pocket with a 
reasonable amount of energy, the fusion process will occur after 
a complete amalgation of the colliding nuclei forming the compound 
nucleus (CN). On the other hand, for sub-barrier energies the 
DNS has not enough energy to pass through the barrier.

In reactions induced by stable beams, the specific role of multi-step 
nucleon-transfers in sub-barrier fusion enhancement still needs to be 
investigated in detail both experimentally  and theoretically
\cite{Pengo83,Stelson88,Rowley92,Timmers98,Zagrebaev03,Stefanini06,Stefanini07,Yang08,Kalkal2010}.
In a complete description of the fusion dynamics the transfer 
channels in standard coupled-channel (CC) calculations 
\cite{Dasgupta,Rowley92,Zagrebaev03,Kalkal2010,CCFULL} have to be 
taken into account accurately. It is known, for instance, that 
neutron transfers may induce a neck region of nuclear matter 
in-between the interacting nuclei favoring the fusion process to 
occur. In this case, neutron pick-up processes can occur when the 
nuclei are close enough to interact each other significantly 
\cite{Stelson88,Rowley92}, if the Q-values of neutron transfers are 
positive. It was shown that sequential neutron transfers can lead to 
the broad distributions characteristic of many experimental fusion 
cross sections. Finite Q-value effects can lead to neutron flow and 
a build up of a neck between the target and projectile \cite{Rowley92}. 
The situation of this neck formation of neutron matter between the 
two colliding nuclei could be considered as a ``doorway state" to 
fusion. In a basic view, this intermediate state induced a barrier 
lowering. As a consequence, it will favor the fusion process at 
sub-barrier energies and enhance significantly the fusion cross 
sections. Experimental results have already shown such enhancement 
of the sub-barrier fusion cross sections due to the neutron-transfer 
channels with positive Q-values \cite{Pengo83,Timmers98}. 

In reactions induced by weakly bound nuclei and/or by halo nuclei, 
the influence on the fusion process of coupling both to collective 
degrees of freedom and to transfer/breakup channels is a key point 
\cite{Liang,Canto,Keeley} for the understanding of N-body systems in 
quantum dynamics \cite{Balantekin}. Due to their very weak binding 
energies, a diffuse cloud of neutrons for $^{6}$He or an extended 
spatial distribution for the loosely bound proton in $^{8}$B would 
lead to larger total reaction (and fusion) cross sections at 
sub-barrier energies as compared to 1D-BPM model predictions. This 
enhancement is well understood in terms of the dynamical processes 
arising from strong couplings to collective inelastic excitations of 
the target (such as "normal" quadrupole and octupole modes) and 
projectile (such as soft dipole resonances). However, in the case of 
reactions where at least one of the colliding nuclei has a 
sufficiently low binding energy for breakup to become a competitive 
process, conflicting conclusions were reported 
\cite{Liang,Canto,Keeley,Beck07a,Beck10}.

Recent studies with Radioactive Ion Beams (RIB) indicate that 
the halo nature of $^{6,8}$He 
\cite{DiPietro04,Navin04,Chatterjee08,Lemasson09,Scuderi11}, 
for instance, does not enhance the fusion 
probability as anticipated. Rather the prominent role of one- and 
two-neutron transfers in $^{6,8}$He induced fusion reactions was 
definitively demonstrated. On the other hand, the effect of 
non-conventional transfer/stripping processes appears to be less 
significant for stable weakly bound projectiles. Several experiments 
involving $^{9}$Be, $^{7}$Li, and $^{6}$Li projectiles on medium-mass 
targets have been undertaken.

\section{Experimental results}
\label{sec:1}

\begin{figure}[t]
\sidecaption[t]
\includegraphics[scale=.55]{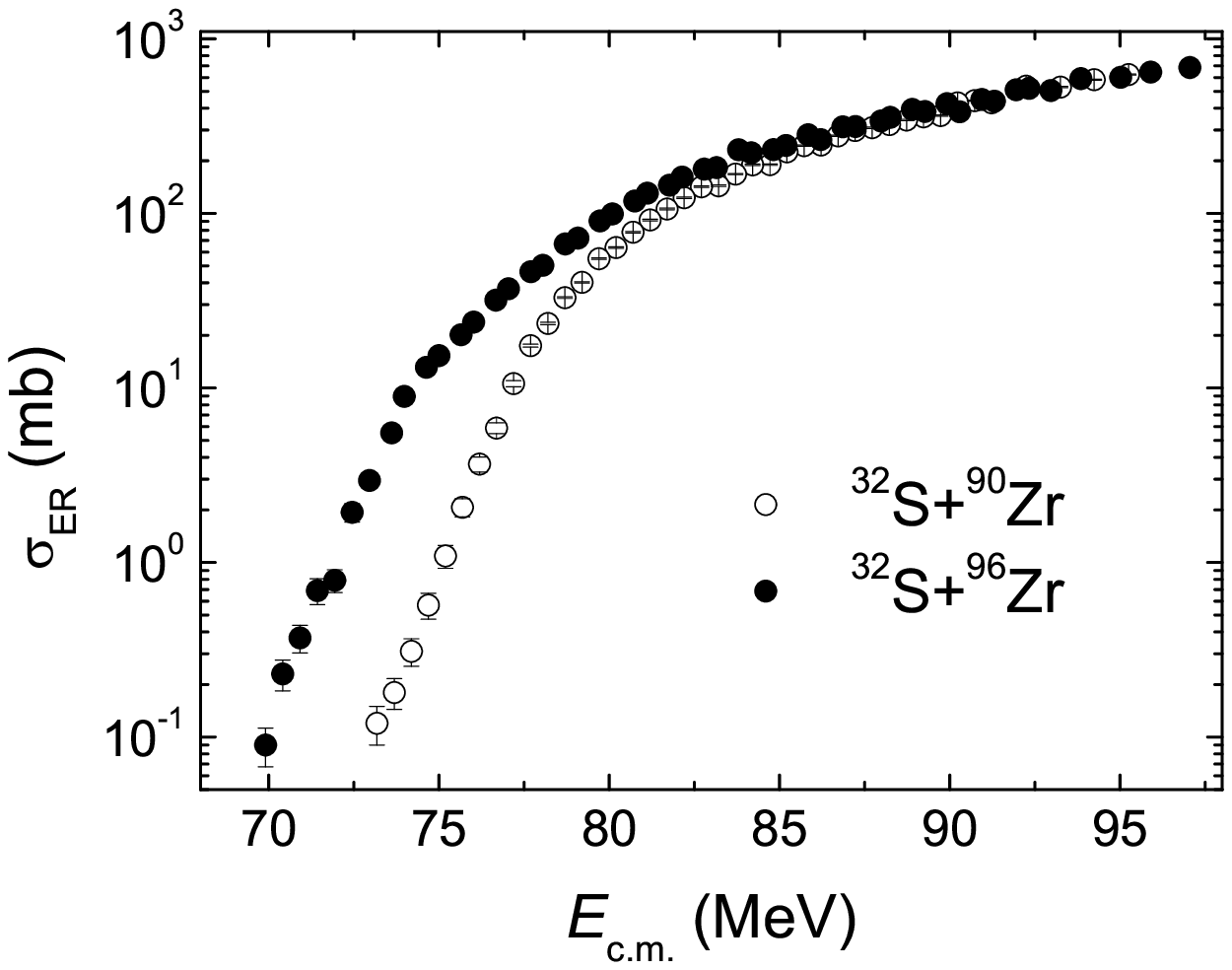}
\caption{Comparison between the fusion-evaporation (ER) excitation 
  functions of $^{32}$S+$^{90}$Zr (open circles) and $^{32}$S+$^{96}$Zr 
  (points) as a function of the center-of-mass energy. The error
  barrs of the experimental data taken from Ref.~\cite{Zhang10} represent
  purely statistics uncertainties. (Courtesy of H.Q. Zhang) 
 }
\label{fig:1}
\end{figure}

\begin{figure}[t]
\sidecaption[t]
\includegraphics[scale=.55]{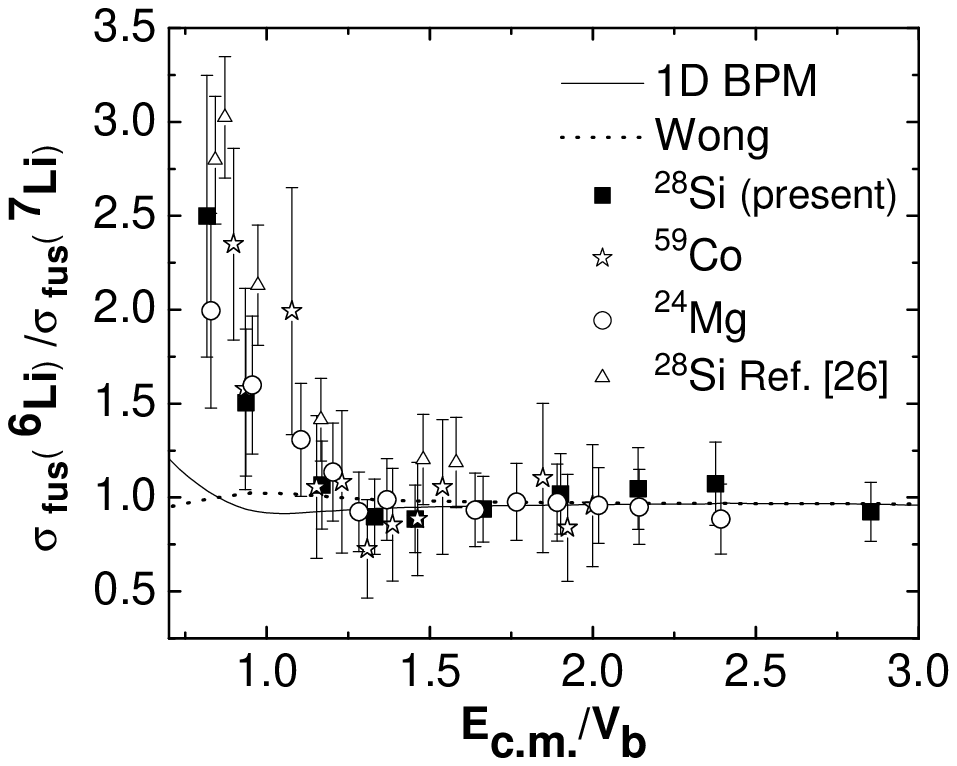}
\caption{Ratios of measured fusion cross sections for $^{6}$Li
and $^{7}$Li projectiles with $^{24}$Mg, $^{28}$Si and 
$^{59}$Co targets as a function of E$_{c.m.}$/V$_b$. The solid line 
gives the 1D-BPM prediction while the dotted line shows results obtained from 
Wong's prescription. (This figure originally shown in Ref.~\cite{Beck03}
for $^{6,7}$Li+$^{59}$Co has been adapted to display comparisons with other 
lighter targets \cite{Sinha08,Ray08,Pakou09,Sinha10}) 
 }
\label{fig:2}
\end{figure}

In this lecture we first present the role of inelastic and transfer channel 
couplings in experimental data obtained in fusion reactions induced by 
stable $^{32}$S projectiles \cite{Zhang10}. The breakup coupling in both  
elastic scattering data and in the fusion data are also shown for
weakly bound $^{6,7}$Li projectiles \cite{Beck03}.

\subsection{$^{32}$S + $^{90}$Zr and $^{32}$S + $^{96}$Zr reactions}

In order to investigate the role of neutron transfers we further study 
$^{32}$S + $^{90}$Zr and $^{32}$S + $^{96}$Zr as benchmark 
reactions. Fig.~1 displays the measured fusion cross sections for 
$^{32}$S + $^{90}$Zr (open circles) and $^{32}$S + $^{96}$Zr 
(points). We present the analysis of excitation functions of evaporation 
residues (ER) cross sections recently measured with high precision (i.e. 
with small energy steps and good statistical accuracy for these 
reactions \cite{Zhang10}).

The differential cross sections of quasi-elastic scattering (QEL) at 
backward angles were previously measured by the CIAE group \cite{Yang08}. 
The analysis of the corresponding BD-QEL barrier distributions (see solid 
points in Fig.~3) already indicated the significant role played by 
neutron tranfers in the fusion processes.

In Fig.~3 we introduce the experimental fusion-barrier (BD-Fusion) 
distributions (see open poins) obtained for the two reactions by using the 
three-point difference method of Ref. \cite{Rowley92} as applied to the data 
points of Ref. \cite{Zhang10} plotted in Fig.~1. It is interesting to note 
that in both cases the BD-Fusion and BD-QEL barrier distributions are almost 
identical up to E$_{c.m.}$ $\approx$ 85 MeV.  

\subsection{$^{6}$Li + $^{59}$Co and $^{7}$Li + $^{59}$Co reactions}

The fusion excitation functions were measured for the $^{6,7}$Li+$^{59}$Co 
reactions \cite{Beck03} at the VIVITRON facility of the IPHC Strasbourg
and the Pelletron facility of Sa\~o Paulo by using $\gamma$-ray techniques.
Their ratios are presented in Fig.~2 with comparisons with other 
lighter targets \cite{Sinha08,Ray08,Pakou09,Sinha10}. The theoretical
curves (1D-BPM \cite{Balantekin, Dasgupta} and Wong \cite{CCFULL} do not take
into account the breakup channel coupling that is discussed in one of the 
following sections in more details.

\section{Coupled channel analysis }
\label{sec:2}

Analysis of experimental fusion cross sections by using standard CC 
calculations is first discussed with the emphasis of the role of 
multi-neutron transfer in the fusion process below the Coulomb barrier 
for $^{32}$S+$^{90,96}$Zr as benchmark reactions.

\subsection{$^{32}$S + $^{90}$Zr and $^{32}$S + $^{96}$Zr reactions}

A new CC computer code named NTFus \cite{NTFus} taking the neutron transfer 
channels into account in the framework of the semiclassical model of 
Zagrebaev \cite{Zagrebaev03} has been developed. The effect of the 
neutron transfer channels yields a fairly good agreement with the data of 
sub-barrier fusion cross sections measured for $^{32}$S + $^{96}$Zr, 
the more neutron-rich reaction~\cite{Zhang10}. This was initially expected 
from the positive Q-values of the neutron transfers as well as from the failure 
of standard CC calculation of quasi-elastic barrier distributions without 
neutron-transfers coupling \cite{Yang08} as shown by the solid line in 
Fig.~3(b). 

By fitting the experimental fusion excitation function displayed in Fig.~1 
with NTFus CC calculation \cite{NTFus}, we concluded \cite{Beck11} that the 
effect of the neutron transfer channels produces significant enhancement of 
the sub-barrier fusion cross sections of $^{32}$S + $^{96}$Zr as compared 
to $^{32}$S + $^{90}$Zr.
A detailed inspection of the $^{32}$S + $^{90}$Zr fusion data presented in 
Fig.~1 along with the negative Q-values of their corresponding neutron 
transfer channels lead us to speculate with the absence of a neutron 
transfer effect on the sub-barrier fusion for this reaction. 
With the semiclassical model developed by Zagrebaev~\cite{Zagrebaev03} we 
propose to definitively demonstrate the 
significant role of neutron transfers for the $^{32}$S + $^{96}$Zr fusion 
reaction by fitting its experimental excitation function with  
\textsc{NTFus} code \cite{NTFus} calculations, as shown in Fig.~3. 

\begin{figure}
\sidecaption[t]
\includegraphics[scale=.56]{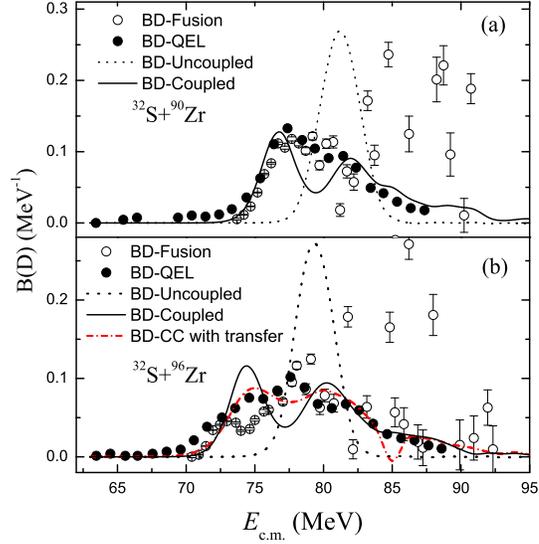}
\caption{Barrier distributions (BD) from the fusion ER (open circles) 
cross sections \cite{Zhang10}, plotted in Fig.1, and quasielastic scattering 
(solid circles) cross sections \cite{Yang08} for $^{32}$S+$^{90}$Zr (a)
and $^{32}$S+$^{96}$Zr (b). The dashed and solid black lines represent
uncoupled calculations (1D-BPM) and the CC calculations without neutron
transfer coupling. The red dash-dotted line represents the CC calculations
with neutron transfer coupling for the $^{32}$S+$^{96}$Zr reaction. (Courtesy
of H.Q. Zhang)}
\label{fig:3}
\end{figure}

The new oriented object \textsc{NTFus} code \cite{NTFus}, using the Zagrebaev 
model \cite{Zagrebaev03} was implemented (at the CIAE) in C++, using the 
compiler of \textsc{ROOT} \cite{ROOT}, following the basic equations
of Ref.~\cite{Zagrebaev04}. 
Let us first remind the values chosen for the deformation parameters and 
the excitation energies that are given in Refs.~\cite{Dasgupta,Raman01,Kebedi05}
(see Tables given in \cite{Beck11} for more details). The quadrupole 
vibrations of both the $^{90}$Zr and $^{96}$Zr are weak in energy; they 
lie at comparable energies.
The $^{96}$Zr nucleus presents a complicated situation~\cite{Corradi11}: 
its low-energy spectrum is dominated by a 2$^{+}$ state at 1.748 MeV and 
by a very collective [B(E3;3$^-$ $\rightarrow$ 0$^+$) = 51 W.u.] 3$^-$ 
state at 1.897 MeV. CC calculations explained the larger sub-barrier
enhancement as due mainly to the strong octupole vibration of the 3$^-$
state in $^{36}$S + $^{96}$Zr \cite{Stefanini00}. 
However, the agreement is not so satisfactory below the barrier 
for $^{32}$S + $^{96}$Zr (see solid line of Fig.~3.b), as well
as  for  $^{40}$Ca + $^{96}$Zr \cite{Timmers98} and, therefore, there is 
the need to take neutron transfers into account.

The main functions of the code \textsc{NTFus} are designed to calculate the 
fusion excitation functions with normalized barrier distribution 
(based on experimental data) given by CCFULL \cite{CCFULL}, we take the 
dynamical deformations into account. In order to introduce the role of
neutron transfers, the \textsc{NTFus} code \cite{NTFus} applies the Zagrebaev 
model~\cite{Zagrebaev03} to calculate the fusion cross sections 
$\sigma_{fus}(E)$ as a function of center-of-mass energy E. Then the fusion 
excitation function can be derived using the following formula
\cite{Zagrebaev03}:\\

T$_{l}$(E)~=
 
\begin{equation}
    \int f(B)\frac{1}{N_{tr}}\sum_{k}\int^{Q_{0}(k)}_{-E}\alpha_{k}(E,l,Q) \\
    \times P_{HW}(B,E+Q,l)dQdB
  \label{eq1}.
\end{equation}

and

\begin{equation}
  \sigma_{fus}(E)=\frac{\pi \hbar^{2}}{2\mu E} \sum^{l_{cr}}_{l=0} (2l+1)T_{l}(E)
\label{eq2}.
\end{equation}

where $T_{l}(E)$ are the transmission coefficients, $E$ is the energy given in 
the center-of-mass system, B and $f(B)$ are the barrier height and the 
normalized barrier distribution function, P$_{HW}$ is the usual Hill-Wheeler
formula. $l$ is the angular momentum whereas 
$l_{cr}$ is the critical angular momentum as calculated by assuming no 
coupling (well above the barrier). 
$\alpha_{k}(E,l,Q)$ and $ Q_{0}(k)$ are, respectively, the probabilities and 
the Q-values for the transfers of $k$ neutrons. And $1/N_{tr}$ is the 
normalization of the total probability taking into account the neutron 
transfers. 

The \textsc{NTFus} code \cite{NTFus} uses the ion-ion potential between two 
deformed nuclei as developped by Zagrebaev and Samarin in 
Ref.~\cite{Zagrebaev04}. Either the standard Woods-Saxon form of the nuclear 
potential or a proximity potential \cite{Blocki77} can be chosen. The code
is also able to predict fusion cross sections for reactions induced by halo 
projectiles \cite{Beck11}; for instance $^{6}$He + $^{64}$Zn 
\cite{Scuderi11,Fisichella11}. In the following, only comparisons for 
$^{32}$S + $^{90}$Zr and $^{32}$S + $^{96}$Zr are discussed.

For the high-energy part of the $^{32}$S + $^{90}$Zr excitation function, 
one can notice a small over-estimation of the fusion cross sections at 
energies above the barrier up to the point used to calculate the critical 
angular momentum. This behavior can be observed at rather high incident 
energies - i.e. between about 82 MeV and 90 MeV (shown as the dashed line 
in Fig. 3.(a) for $^{32}$S + $^{90}$Zr reaction). We want to stress that 
the corrections do not affect our conclusions that the transfer channels 
have a predominant role below the barrier for $^{32}$S + $^{96}$Zr reaction, 
as shown by the dotted-dashed red curve in Fig.~3.(b). 

As expected, we obtain a good agreement with calculations not taking any 
neutron transfer coupling into account for $^{32}$S + $^{90}$Zr as shown
by the solid line of Fig.~3.(a) (the dashed line are the results of 
calculations performed without any coupling). On the other hand, there is no 
significant over-estimation at sub-barrier energies. As a consequence, it 
is possible to observe the strong effect of neutron transfers on the 
fusion for the $^{32}$S + $^{96}$Zr reaction at sub-barrier energies. 
Moreover, the barrier distribution function $f(B)$ extracted from the data 
contains the information of the neutron transfers. These information are also 
contained in the transmission coefficients, which are the most 
important parameters for the fusion cross sections to be calculated 
accurately. The $f(B)$ function as calculated with the three-point formula 
\cite{Rowley92} will mimic the differences induced by the neutron transfer 
taking place in sub-barrier energies where the cross section variations are 
very small (only visible if a logarithm scale is employed for the fusion 
excitation function). It is interesting to note that the Zagrebaev 
model \cite{Zagrebaev03} implies a modification of the Hill-Wheeler 
probability and does not concern the barrier distribution function $f(B)$.
Finally, the code allows us to perform each calculation by taking the 
neutron transfers into account or not.

The calculation with the neutron transfer effect is performed up to the 
channel +4n (k=4), but we have seen that we obtain the same overall agreement
with data up to channels +5n and +6n \cite{Beck11}. 
As we can see on Fig. 3.(b), the solid line representing standard CC calculations
without the neutron transfer coupling (the dotted line is given for uncoupled
calculations) does not fit the experimental data 
well at sub-barrier energies. On the other hand, the dotted line displaying 
NTFus calculations taking the neutron transfer coupling into account agrees
perfectly well with the data. As expected, the Zagrebaev semiclassical model's 
correction applied at sub-barrier energies enhances the calcutated cross 
sections. Moreover, it allows to fit the data reasonably well and therefore 
illustrates the strong effect of neutron transfers for the fusion of 
$^{32}$S + $^{96}$Zr at subbarrier energies. 
 
The present full CC analysis of $^{32}$S + $^{96}$Zr fusion data 
\cite{Zhang10,Beck11} using NTFus \cite{NTFus} confirms perfectly well first 
previous CC calculations \cite{Zagrebaev03} describing well the earlier 
$^{40}$Ca + $^{90,96}$Zr fusion data \cite{Timmers98} and, secondly, very 
recent fragment-$\gamma$ coincidences measured for $^{40}$Ca + $^{96}$Zr 
multi-neutron transfer channels \cite{Corradi11}.

\subsection{$^{6}$Li + $^{59}$Co and $^{7}$Li + $^{59}$Co reactions}

\begin{figure}
\sidecaption[t]
\includegraphics[scale=.48]{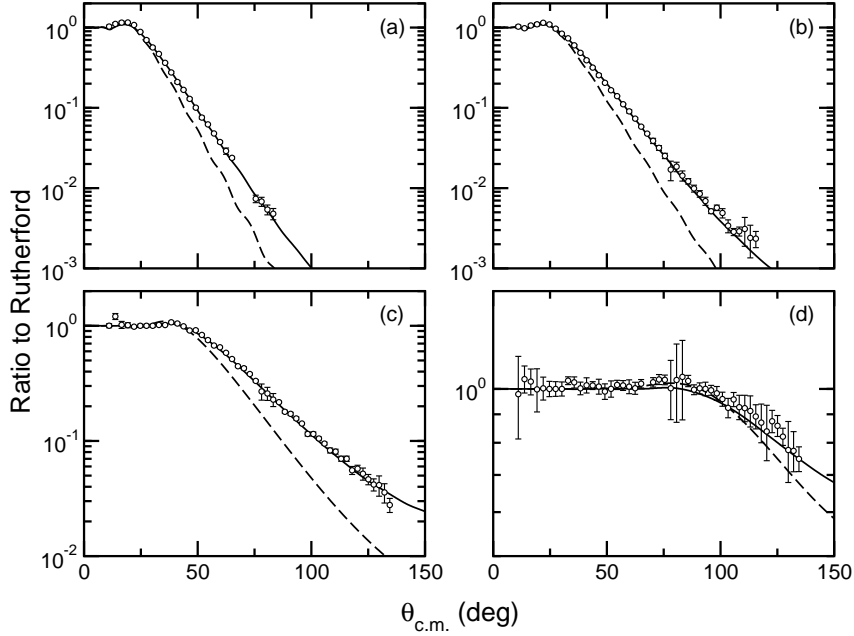}
\caption{Ratios of the elastic scattering cross-sections to the
Rutherford cross sections as a function of c.m. angle for the $^{6}$Li+$^{59}$Co
system~\cite{Beck07b}. The curves correspond to CDCC calculations with
(solid lines) and without (dashed lines) $^{6}$Li $\rightarrow$ $\alpha$ + $d$
breakup couplings to the continuum for incident $^6$Li energies of (a) 30 MeV,
(b) 26 MeV, (c) 18 MeV and (d) 12 MeV. (This figure has been adapted
from the work of Ref.~\cite{Beck07b})}
\label{fig:4}
\end{figure}

For reactions induced by weakly bound nuclei 
\cite{Beck07a,Beck10,Sinha08,Ray08,Pakou09,Sinha10,Souza09,Souza10a,Souza10b}
and exotic nuclei 
\cite{DiPietro04,Navin04,Chatterjee08,Lemasson09,Scuderi11,Fisichella11,
DiPietro10,Mazzocco10,Kohley11,Aguilera12,Rudolph12},
the breakup channel is open and plays a key role in the fusion process near 
the Coulomb barrier similarly to the transfer-channel coupling described in 
the previous section. It is therefore appropriate to use the 
Continuum-Discretized Coupled-Channel (CDCC) approach 
\cite{Beck07b,Keeley10,Diaz03,Beck11b} to describe the influence of the
breakup channel in both the elastic scattering and the fusion process at
sub-barrier energies.

Theoretical calculations (including CDCC predictions given in 
Refs.~\cite{Beck07b,Diaz03} indicate only a small enhancement of total fusion 
for the more weakly bound $^{6}$Li below the Coulomb barrier (see curves of 
Fig.2), with similar cross sections for both $^{6,7}$Li+$^{59}$Co reactions 
at and above the barrier \cite{Beck03}. It is interesting to notice, however, 
that the same conclusions have been reached for other targets such
as $^{24}$Mg \cite{Ray08} and $^{28}$Si \cite{Sinha08,Pakou09,Sinha10} as 
can be clearly seen in the plot of Fig.~2. Thess results are consistent with 
rather low breakup cross sections measured for the 
$^{6,7}$Li+$^{59}$Co reactions even at incident energies larger than the Coulomb 
barrier \cite{Souza09,Souza10a,Souza10b}. But the coupling of the breakup 
channel is extremely important for the CDCC analysis of the angular 
distributions of the elastic scattering \cite{Beck07b} as shown
in Fig.~4 for $^{6}$Li+$^{59}$Co. The curves show the results of calculations 
with (solid lines) and without (dashed lines) $^{6,7}$Li $\rightarrow$ $\alpha$ 
+ $d,t$ breakup couplings. The main conclusion is that effect of breakup on 
the elastic scattering is stronger for $^{6}$Li than $^{7}$Li.

\begin{figure}[t]
\sidecaption[t]
\includegraphics[scale=.65]{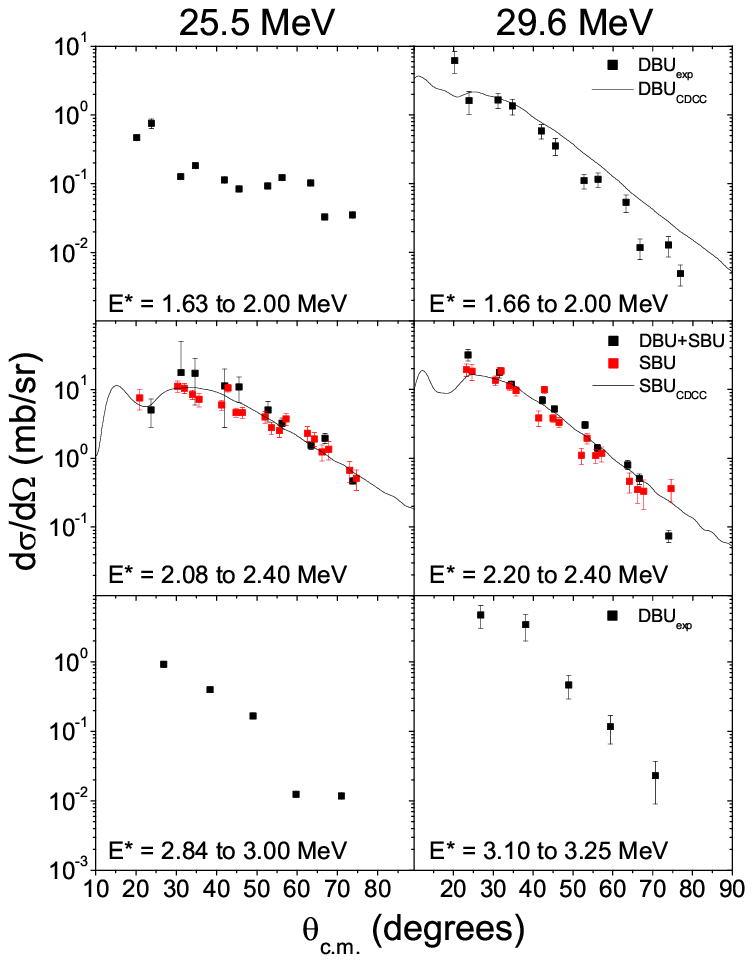}
\caption{Experimental \cite{Souza09,Souza10a,Souza10b} and theoretical 
CDCC \cite{Beck11} angular distributions
for the SBU and DBU projectile breakup processes (see text for 
details) obtained at E$_{lab}$ = 25.5 MeV and 29.6 MeV for 
$^{6}$Li+$^{59}$Co. The chosen experimental continuum excitation energy ranges 
are given. (Courtesy of F.A, Souza)
 }
\label{fig:5}

\end{figure}
\begin{figure}
\sidecaption[t]
\includegraphics[scale=.48]{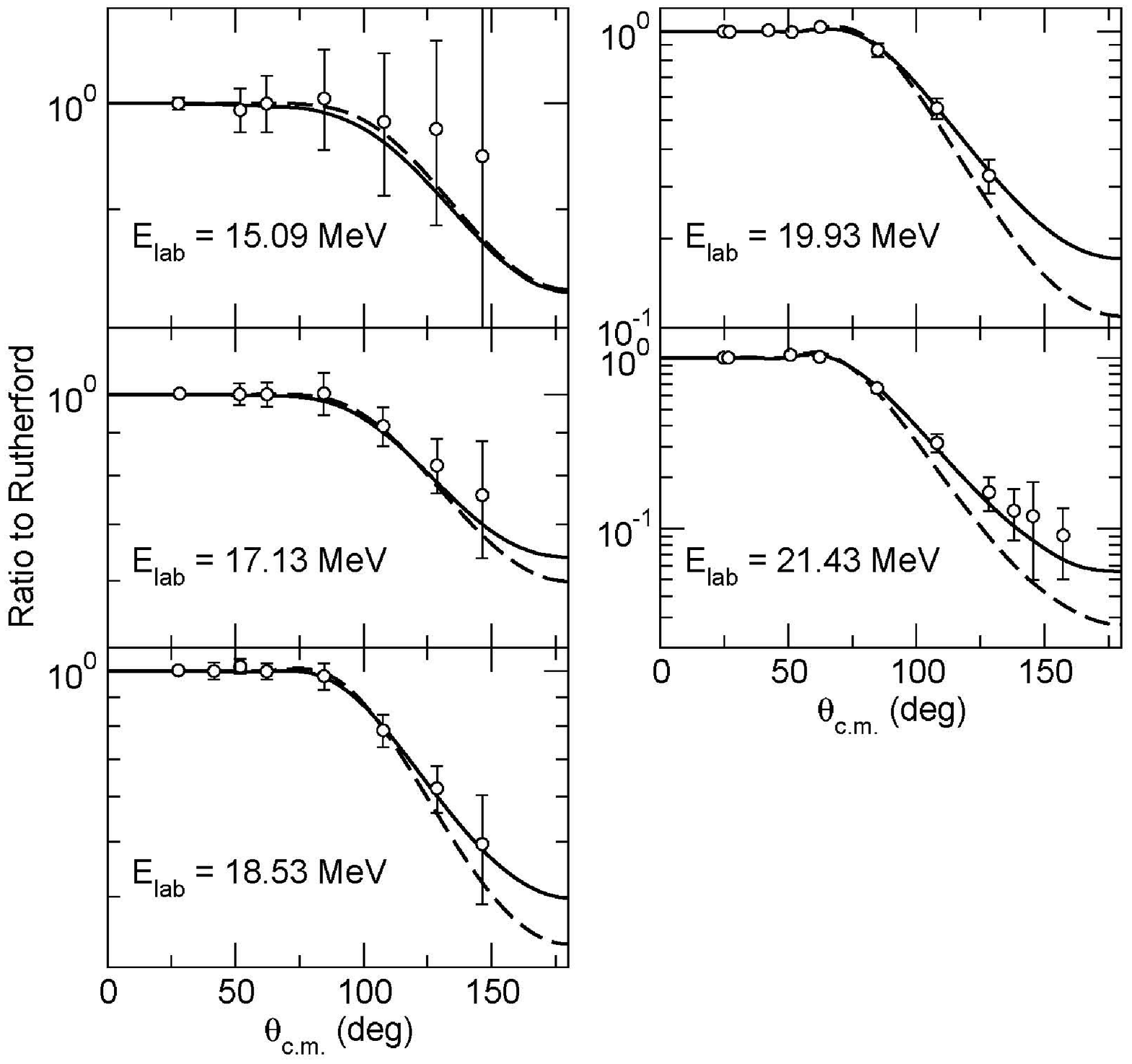}
\caption{Ratios of the elastic scattering cross-sections to the
Rutherford cross sections as a function of c.m. angle for the $^{7}$Be+$^{58}$Ni
system~\cite{Aguilera09} for incident $^7$Be energies of 
(a) 15.09 MeV, (b) 17.13 MeV, (c) 18.53 MeV (d) 19.93 MeV and (e) 21.43 MeV. 
The solid and dashhed curves denote full and no
coupling to the continuum.
(This figure has been adapted from the work of Ref.~\cite{Beck10})}
\label{fig:6}
\end{figure}

A more detailed investigation of the breakup process in the $^{6}$Li+$^{59}$Co 
reaction with particle coincidence techniques is now proposed to discuss 
the interplay of fusion and breakup processes. Coincidence data compared to 
three-body kinematics calculations reveal a way how to disentangle the 
contributions of breakup, incomplete fusion and/or transfer-reemission 
processes \cite{Souza09,Souza10a,Souza10b}. 

Fig.~5 displays experimental (full rectangles) and theoretical angular 
distributions (solid lines) for the sequential (SBU) and direct (DBU) 
projectile breakup processes at the two indicated bombarding energies 
for the $^{6}$Li+$^{59}$Co reaction. In the CDCC calculations the 
$\alpha$ + $d$ binning scheme is appropriately altered to accord exactly 
with the measured continuum excitation energy ranges. For this reaction it  
was not necessary to use a sophisticated four-body CDCC framework. 
The CDCC cross sections \cite{Beck07b} are in agreement with the experimental 
ones \cite{Beck07a,Souza10a,Souza10b}, both in shapes and magnitudes within 
the uncertainties. The relative contributions of the $^{6}$Li SBU and DBU to 
the incomplete fusion/transfer process has been discussed in great details 
in Refs.~\cite{Souza09,Souza10a,Souza10b} by considering the corresponding 
lifetimes obtained by using a semi-classical approach fully described
in a previous publication \cite{Souza09}. We conclude  that the flux 
diverted from complete fusion to incomplete fusion would arise essentially 
from DBU processes via high-lying continuum (non-resonant) states of 
$^{6}$Li; this is due to the fact that both the SBU mechanism and the 
low-lying DBU processes from low-lying resonant $^{6}$Li states occur at 
large internuclear distances \cite{Souza09,Souza10a,Souza10b}. Work is in 
progress to study incomplete fusion for $^{6}$Li+$^{59}$Co within a newly 
developed 3-dimensional classical trajectory model \cite{Diaz07}.

\subsection{Coupled-channel calculations for reactions induced by halo nuclei}

As far as exotic halo projectiles are concerned we have initiated a systematic
study of $^{8}$B and $^{7}$Be induced reactions data \cite{Aguilera09} with 
an improved CDCC method \cite{Keeley10}. Fig.~6 displays the analysis of
the elastic scattering for the $^{7}$Be+$^{58}$Ni system~\cite{Aguilera09}.
The curves correspond to CDCC calculations with
(solid lines) and without (dashed lines) $^{7}$Be $\rightarrow$ $\alpha$ +
$^3$He breakup couplings to the continuum. The $^6$Li and  $^{7}$Be
calculations were similar, but with a finer continuum binning for $^{7}$Be.
As compared to $^{7}$Be+$^{58}$Ni (similar to $^{6,7}$Li+$^{58,64}$Ni) the 
CDCC analysis of $^{8}$B+$^{58}$Ni reaction \cite{Keeley10} while exhibiting 
a large breakup cross section (consistent with the systematics) is rather 
surprizing as regards the consequent weak coupling effect found to be 
particularly small on the near-barrier elastic scattering. 

Recently, the scattering process of $^{17}$F from $^{58}$Ni target was
investigated \cite{Mazzocco10} slightly above the Coulomb barrier
and total reaction cross sections were extracted from the Optical-Model
analysis. The small enhancement as compared to the reference (tightly
bound) system $^{16}$O+$^{58}$Ni is here related to the low binding energy
of the $^{17}$F valence proton. This moderate effect is mainly triggered
from a transfer effect, as observed for the 2$n$-halo $^{6}$He
\cite{DiPietro04,Navin04} and the 1$n$-halo $^{11}$Be \cite{DiPietro10} 
in contrast to the 1$p$-halo $^{8}$B+$^{58}$Ni reaction where strong 
enhancements are trigerred from a breakup process \cite{Aguilera12}.

\newpage

\section{Summary, conclusions and outlook}

We have investigated the fusion process (excitation functions and extracted 
barrier distributions \cite{Zhang10}) at near- and sub-barrier energies for 
the two neighbouring reactions $^{32}$S + $^{90}$Zr and $^{32}$S + $^{96}$Zr. 
For this purpose a new computer code named NTFus \cite{NTFus} has been 
developped by taking the coupling of the multi-neutron transfer channels 
into account by using the semiclassical model of Zagrebaev \cite{Zagrebaev03}. 

The effect of neutron couplings provides a fair agreement with the present 
data of sub-barrier fusion for $^{32}$S + $^{96}$Zr. This was initially 
expected from the positive Q-values of the neutron transfers as well as from 
the failure of previous CC calculation of quasi-elastic barrier distributions 
without coupling of the neutron transfers \cite{Yang08}. With the agreement 
obtained by fitting the present experimental fusion excitation function
and the CC calculation at sub-barrier energies, we conclude that the effect 
of the neutron transfers produces a rather significant enhancement of the 
sub-barrier fusion cross sections of $^{32}$S + $^{96}$Zr as compared to 
$^{32}$S + $^{90}$Zr. At this point we did not try to reproduce the details 
of the fine structures observed in the fusion barrier distributions. We 
believe that to achieve this final goal it will first be necessary to measure 
the neutron transfer cross sections to provide more information on the 
coupling strength of neutron transfer because its connection with fusion
is not yet fully understood \cite{Corradi11}. 

In the second part of this lecture, we have studied the breakup coupling
on elastic scattering and fusion by using the CDCC approach with a 
particular emphasis on a very detailed analysis of the $^{6}$Li+$^{59}$Co 
reaction. The CDCC formalism, with continuum--continuum couplings taken 
into account, is probably one of the most reliable methods available 
nowadays to study reactions induced by exotic halo nuclei, although many 
of them have added complications like core excitation and three-body 
structure. The respective effects of transfer/breakup are finally outlined 
for reactions induced by 1$p$-halo, 1$n$-halo and 2$n$-halo nuclei.

The complexity of such reactions, where many processes compete on an equal 
footing, necessitates kinematically and spectroscopically complete 
measurements \cite{Papka12}, i.e.\ ones in which all processes from elastic 
scattering to fusion are measured simultaneously, providing a technical 
challenge in the design of broad range detection systems. A full 
understanding of the reaction dynamics involving couplings to the breakup 
and nucleon-transfer channels will need high-intensity RIB and precise 
measurements of elastic scattering, fusion and yields leading to the 
breakup itself. A new experimental program with SPIRAL beams and 
medium-mass targets is getting underway at GANIL. 

\begin{acknowledgement}

I would like to thank A. Diaz-Torres, N. Keeley, F.A. Souza and A. Richard
for very fruitfull discussions on many theoretical aspects of this lecture.
 
\end{acknowledgement}

\newpage


\begin{thebibliography}{10}
\providecommand{\url}[1]{{#1}}
\providecommand{\urlprefix}{URL }
\expandafter\ifx\csname urlstyle\endcsname\relax
  \providecommand{\doi}[1]{DOI~\discretionary{}{}{}#1}\else
  \providecommand{\doi}{DOI~\discretionary{}{}{}\begingroup
  \urlstyle{rm}\Url}\fi
  
\bibitem{Balantekin} A.B. Balantekin and N. Takigawa, Rev. Mod. Phys. 
\bf 70\rm, 77 (1998); arXiv:\bf nucl-th/9708036 \rm (1997).
\bibitem{Dasgupta} M. Dasgupta, D.J. Hinde, N. Rowley, and A.M. Stefanini, 
Annu. Rev. Nucl. Part. Sci. \bf 48\rm, 401 (1998).
\bibitem{Liang} J.F. Liang and C. Signorini, Int. J. Mod. Phys. E 
\bf 14\rm, 1121 (2005); \rm arXiv:\bf nucl-ex/0504030 \rm (2005).
\bibitem{Canto} L.F. Canto, P.R.S. Gomes, R. Donangelo, and 
M.S. Hussein, Phys. Rep. \bf 424\rm, 1 (2006).
\bibitem{Keeley} N. Keeley, R. Raabe, N. Alamanos, and J.-L. Sida, 
Prog. Part. Nucl. Phys. \bf 59\rm 579 (2007); arXiv:\bf nucl-th/0702038 \rm (2007).
\bibitem{Pengo83} R. Pengo, D. Evers, K.E.G. Lobner, U. Quade, K. Rudolph, 
S.J. Skorka, and I. Weidl, Nucl. Phys. A \bf 411\rm, 256 (1983).
\bibitem{Stelson88} P.H. Stelson, Phys. Lett. B \bf 205\rm, 190 (1988).
\bibitem{Rowley92} N. Rowley, G.J. Thompson, and M.A. Nagarajan, Phys. Lett.
B \bf 282\rm, 276 (1992). 
\bibitem{Timmers98} H. Timmers, D. Ackermann, S. Beghini, L. Corradi, 
J.H. He, G. Montagnoli, F. Scarlassara, A.M. Stefanini, and N. Rowley, 
Nucl. Phys. \bf A633\rm, 421 (1998).
\bibitem{Zagrebaev03} V.I. Zagrebaev, Phys. Rev. C \bf 67\rm, 061601 (2003).
\bibitem{Stefanini06} A.M. Stefanini, F. Scarlassara, S. Beghini, G. Montagnoli, 
R. Silvestri, M. Trotta, B.R. Behera, L. Corradi, E. Fioretto, A. Gadea, Y.W. Wu, 
S. Szilner, H.Q. Zhang, Z.H. Liu, M. Ruan, F. Yang, and N. Rowley,
Phys. Rev. C \bf 73\rm, 034606 (2006).
\bibitem{Stefanini07} A.M. Stefanini, B.R. Behera, S. Beghini, L. Corradi, 
E. Fioretto, A. Gadea, G. Montagnoli, N. Rowley, F. Scarlassara, S. Szilner, 
and M. Trotta, Phys. Rev. C \bf 76\rm, 014610 (2007).
\bibitem{Yang08} F. Yang, C.J. Lin, X.K. Wu, H.Q. Zhang, C.L. Zhang, 
P. Zhou, and Z.H. Liu, Phys. Rev. C \bf 77\rm, 014601 (2008).
\bibitem{Kalkal2010} Sunil Kalkal, S. Mandal, N. Madhavan, E. Prasad, Shashi 
Verma, A. Jhingan, Rohit Sandal, S. Nath, J. Gehlot, B.R. Behera, Mansi 
Saxena, Savi Goyal, Davinder Siwal, Ritika Garg, U.D. Pramanik, Suresh Kumar, 
T. Varughese, K.S. Golda, S. Muralithar, A.K. Sinha, and R. Singh, Phys. 
Rev. C \bf 81\rm, 044610 (2010).
\bibitem{CCFULL} K. Hagino, N. Rowley, and A.T. Kruppa, Comput. Phys.
Commun. \bf 123\rm, 143 (1999); arXiv:\bf nucl-th/9903074 \rm (1999). 
\bibitem{Beck07a} C.~Beck, Nucl. Phys. A {\bf 787}, 251 (2007); 
arXiv:\bf nucl-ex/0701073 \rm (2007); arXiv:\bf nucl-th/0610004 \rm (2006).
\bibitem{Beck10} C.~Beck, N. Rowley, P. Papka, S. Courtin, M. Rousseau, 
F.A. Souza, N. Carlin, R. Liguori Neto, M.M. de Moura, M.G. Del Santo, 
A.A.P. Suaide, M.G. Munhoz, E.M. Szanto, A. Szanto de Toledo, N. Keeley, 
A. Diaz-Torres, and K. Hagino, Nucl. Phys. A {\bf 834}, 440 (2010);
arXiv:\bf 0910.1672 \rm (2010).
\bibitem{DiPietro04} A.~Di~Pietro, P. Figuera, F. Amorini, C. Angulo, G.
Cardella, S. Cherubini, T. Davinson, D. Leanza, J. Lu, H. Mahmud, M. Milin,
A. Musumarra, A. Ninane, M. Papa, M.G. Pellegriti, R. Raabe, F. Rizzo,
C. Ruiz, A.C. Shotter, N. Soic, and S. Tudisco, Phys. Rev. C {\bf 69}, 
044601 (2004).
\bibitem{Navin04} A. Navin, V. Tripathi, Y. Blumenfeld, V. Nanal, C. Simenel,
J.M. Casandjian, G. de France, R. Raabe, D. Bazin, A. Chatterjee, M. Dasgupta,
S. Kailas, R.C. Lemmon, K. Mahata, R.G. Pilay, E.C. Pollacco, K. Ramachandran,
M. Rejmund, A Shrivastava, J.L. Sida, and E. Tryggestad, Phys. Rev. C 
{\bf 70}, 044601 (2004).
\bibitem{Chatterjee08} A.~Chatterjee, A. Navin, A. Shrivastava, 
S. Bhattacharyya, M. Rejmund, N. Keeley, V. Nanal, J. Nyberg, R. G. Pillay,
K. Ramachandran, I. Stefan, D. Bazin, D. Beaumel, Y. Blumenfeld, 
G. de France, D. Gupta, M. Labiche, A. Lemasson, R. Lemmon, R. Raabe, 
J. A. Scarpaci, C. Simenel, and C. Timis, Phys. Rev. Lett. {\bf 101}, 
032701 (2008).
\bibitem{Lemasson09} A.~Lemasson, Shrivastava, A. Navin, M. Rejmund, 
N. Keeley, V. Zelevinsky, S. Bhattacharyya, A. Chatterjee, G. de France,
B. Jacquot, V. Nanal, R.G. Pillay, R. Raabe, and C. Schmitt, Phys. Rev. Lett. 
{\bf 103}, 232701 (2009).
\bibitem{Scuderi11} V. Scuderi, A. Di Pietro, P. Figuera, M. Fisichella,  
F. Amorini, C. Angulo, G. Cardella, E. Casarejos, M. Lattuada, M. Milin,  
A. Musumarra, M. Papa, M. G. Pellegriti, R. Raabe, F. Rizzo, N. Skukan, 
D. Torresi, and M. Zadro, Phys. Rev. C \bf 84\rm, 064604 (2011).
\bibitem{Zhang10} H.Q. Zhang, C.J. Lin, F. Yang, H.M. Jia, X.X. Xu, F. Jia, 
Z.D. Wu, S.T. Zhang, Z.H. Liu, A. Richard, C. Beck, Phys. Rev. C 
\bf 82\rm, 054609 (2010); arXiv: \bf 1005.0727 \rm (2010).  
\bibitem{Beck03} C. Beck, F.A. Souza, N. Rowley, S.J. Sanders,
N. Aissaoui, E.E. Alonso, P. Bednarczyk, N. Carlin, S. Courtin,
A. Diaz-Torres, A. Dummer, F. Haas, A. Hachem, K. Hagino, F.
Hoellinger, R.V.F. Janssens, N. Kintz, R. Liguori Neto, E. Martin, 
M.M. Moura, M.G. Munhoz, P. Papka, M. Rousseau, A. Sanchez i Zafra, 
O. Stezowski, A.A. Suaide, E.M. Szanto, A. Szanto de Toledo, S.
Szilner, and J.Takahashi, Phys. Rev. C \bf 67\rm, 054602 (2003);
\rm arXiv:nucl-ex/\bf 0411002\rm (2004).
\bibitem{Sinha08} M.~Sinha, H. Majumdar, P. Basu, Subinit Roy, R. 
Bhattacharya, M. Biswas, M.K. Pradhan, and S. Kailas, Phys. Rev. C {\bf 78}, 
027601 (2008); arXiv:\bf nucl-ex:0805.0953 \rm (2008).
\bibitem{Ray08} M.~Ray, A. Mukherjee, M.K. Pradhan, Ritesh Kshetri, 
M. Saha Sarkar, R. Palit, I. Majumdar, P.K. Joshi, H.C. Jain, and 
B. Dasmahapatra, Phys.~Rev.~C {\bf 78}, 064617 (2008); 
arXiv:\bf nucl-ex:0805.0953 \rm (2008).
\bibitem{Pakou09} A.~Pakou, K. Rusek, N. Alamanos, X. Aslanoglou, M. 
Kokkoris, A. Lagoyannis, T.J. Mertzimekis, A. Musumarra, N.G. Nicolis, D. 
Pierroutsakou, and D. Roubos, Eur. Phys. J. A {\bf 39}, 187 (2009).
\bibitem{Sinha10} M.~Sinha, H. Majumdar, P. Basu, Su. Roy, R. Bhattacharya, 
M. Biswas, M.K. Pradhan, R. Palit, I. Mazumdar, and S. Kailas, Eur. Phys. 
J. A {\bf 44}, 403 (2010).
\bibitem{NTFus} H.Q. Zhang, H.Q. Zhang, C.L. Zhang, H.M. Jia, C.J. Lin, 
F. Yang, Z.H. Liu, Z.D. Wu, F. Jia, X.X. Xu, A. Richard, A.K. Nasirov, 
G. Mandaglio, M. Manganaro, G. Giardina, and K. Hagino, AIP Conf. Proc.
\bf 1235\rm, 50 (2010). 
\bibitem{Beck11} A. Richard, C. Beck, and H.Q. Zhang, EPJ Web of 
Conferences \bf 17\rm, 08005 (2011); arXiv:\bf 1104.5333 \rm (2011).
\bibitem{ROOT} http://root.cern.ch, website of ROOT.
\bibitem{Zagrebaev04} V.I. Zagrebaev and V.V. Samarin, Physics of Atomic 
Nuclei \bf 67 \rm No.8, 1462 (2004).
\bibitem{Raman01} S. Raman, C.W. Nestor, and P. Tikkanen, At. Data Nucl. Data
Tables \bf 78\rm, 1 (2001).
\bibitem{Kebedi05} T. Kebedi and R.H. Spear. At. Data Nucl. Data
Tables \bf 89\rm, 77 (2005).
\bibitem{Corradi11} L. Corradi, S. Szilner, G. Pollarolo, G. Colo, P. Mason, 
E. Farnea, E. Fioretto, A. Gadea, F. Haas, D. Jelavic-Malenica, N. Marginean, 
C. Michelagnoli, G. Montagnoli, D. Montanari, F. Scarlassara, N. Soic, 
A.M. Stefanini, C.A. Ur, J.J. Valiente-Dobon, Phys. Rev. C {\bf 84}, 034603 
(2011).
\bibitem{Stefanini00} A.M. Stefanini, L. Corradi, A.M. Vinodkumar, F. Yang, 
F. Scarlassara, G. Montagnoli, S. Beghini, M. Bisogno, Phys. Rev. C \bf 62\rm, 
014601 (2000).
\bibitem{Blocki77} J. Blocki, J. Randrup, W.J. Swiatecki, and C. F. Tsang, 
Annals of Physics \bf 105\rm, 427 (1977).
\bibitem{Fisichella11} M. Fisichella, V. Scuderi, A. Di Pietro, P. Figuera, 
M. Lattuada, C. Marchetta, M. Milin, A. Musumarra, M.G. Pellegriti, N. Skukan, 
E. Strano, D. Torresi, and M. Zadro, J. Phys.: Conf. Ser. \bf 282\rm, 012014 
(2011).
\bibitem{Souza09} F.A.~Souza, C. Beck, N. Carlin, N. Keeley, R. Liguori Neto, 
M.M. de Moura, M.G. Munhoz, M.G. Del Santo, A.A.P. Suaide, E.M. Szanto, and
A.Szanto de Toledo, Nucl. Phys. A {\bf 821}, 36 (2009); arXiv:\bf 0811.4556 
\rm (2008).
\bibitem{Souza10a} F.A.~Souza, N. Carlin, C. Beck, N. Keeley, A. 
Diaz-Torres, R. Liguori Neto, C. Siqueira-Mello, M.M. de Moura, M.G. Munhoz, 
R.A.N. Oliveira, M.G. Del Santo, A.A.P. Suaide, E.M. Szanto, and 
A. Szanto de Toledo, Nucl. Phys. A {\bf 834}, 420 (2010); 
arXiv:\bf 0910.0342 \rm (2010). 
\bibitem{Souza10b} F.A.~Souza, N. Carlin, C. Beck, N. Keeley, A. Diaz-Torres, 
R. Liguori Neto, C. Siqueira-Mello, M.M. de Moura, M.G. Munhoz, R.A.N. 
Oliveira, M.G. Del Santo, A.A.P. Suaide, E.M. Szanto, and A. Szanto de 
Toledo, Eur. Phys. J. A {\bf 44}, 181 (2010); arXiv:\bf 0909.5556 \rm (2009).
\bibitem{DiPietro10} A.~Di~Pietro, G. Randisi, V. Scuderi, L. Acosta, 
F. Amorini, M.J.G. Borge, P. Figuera, M. Fisichella, L.M. Fraile, 
J. Gomez-Camacho, H. Jeppesen, M. Lattuada, I. Martel, M. Milin, A. Musumarra, 
M. Papa, M.G. Pellegriti, F. Perez-Bernal, R. Raabe, F. Rizzo, D. Santonocito, 
G. Scalia, O. Tengblad, D. Torresi, A.M. Vidal, D. Voulot, F. Wenander, and
M. Zadro, Phys. Rev. Lett. {\bf 105}, 022701 (2010).
\bibitem{Mazzocco10} M. Mazzocco, C. Signorini, D. Pierroutsakou,
T. Glodariu, C. Boiano, F. Farinon, A. Di Pietro, P. Figuera,
D. Filipescu, L. Fortunato, A. Guglielmetti, G. Inglima, M. La Commara, 
M. Lattuada, P. Lotti, C. Mazzocchi, P. Molini, A. Musumarra, A. Pakou, 
C. Parascandolo, N. Patronis, M. Romoli, M. Sandoli, V. Scuderi,  
F. Soramel, L. Stroe, D. Torresi, E. Vardaci, and A. Vitturi, 
Phys. Rev. C \bf 82\rm, 054604 (2010).
\bibitem{Kohley11} Z. Kohley, F. Liang, D. Shapira, R.L. Varner, C.J. 
Gross, J.M. Allmond, A.L. Caraley, E.A. Coello, F. Favela, K. Lagergren, 
and P.E. Mueller, Phys. Rev. Lett. \bf 107\rm, 202701 (2011).
\bibitem{Aguilera12} E.F. Aguilera and J.J. Kolata, Phys. Rev. C \bf 85\rm, 
014603 (2012).
\bibitem{Rudolph12} M.J. Rudolph, Z.Q. Gosser, K. Brown, S. Hudan, R.T. 
de Souza, A. Chbihi, B. Jacquot, M. Famiano, J.F. Liang, D. Shapira, and 
D. Mercier, Phys. Rev. C \bf 85\rm, 024605 (2012).
\bibitem{Beck07b} C.~Beck, N.~Keeley, and A.~Diaz-Torres, Phys. Rev. C {\bf 75}, 
054605 (2007); arXiv:\bf nucl-th/0703085 \rm (2007).
\bibitem{Keeley10} N.~Keeley, R.S.~Mackintosh, and C.~Beck, Nucl. Phys. A {\bf 834}, 
792 (2010). 
\bibitem{Diaz03} A.~Diaz-Torres, I.J.~Thompson, C.~Beck, Phys.~Rev.~C {\bf 68}, 
044607 (2003); arXiv:\bf nucl-th/0307021 \rm (2003).
\bibitem{Beck11b} C. Beck, N. Rowley, P. Papka, S. Courtin, M. Rousseau, 
F.A. Souza, N. Carlin, F. Liguori Neto, M.M. De Moura, M.G. Del Santo, 
A.A.I. Suade, M.G. Munhoz, E.M. Szanto, A. Szanto De Toledo, N. Keeley, 
A. Diaz-Torres, and K. Hagino, Int. J. Mod. Phys. \bf E20\rm, 943 (2011);
arXiv:\bf 1009.1719 \rm (2010). 
\bibitem{Diaz07} A. Diaz-Torres, D.J. Hinde, J.A. Tostevin, M. Dasgupta, 
L.R. Gasques, Phys. Rev. Lett. \bf 98\rm, 152701 (2007); 
arXiv:\bf nucl-th/0703041 \rm (2007). 
\bibitem{Aguilera09} E.F. Aguilera, E.F. Aguilera, E. Martinez-Quiroz, D. Lizcano, 
A. Gomez-Camacho, J.J. Kolata, L.O. Lamm, V. Guimaraes, R. Lichtenthaler, 
O. Camargo, F.D. Becchetti, H. Jiang, P.A. DeYoung, P.J. Mears, and
T.L.Belyaeva, Phys. Rev. C \bf 79\rm, 021601 (2009).
\bibitem{Papka12} P. Papka and C. Beck, Clusters in Nuclei, Vol. 2 (Ed.) C. Beck,
Lecture Notes in Physics \bf 848\rm, 299 (2012).

\end{thebibliography}

\end{document}